# Connecting the Dots: Leveraging Social Network Analysis to Understand and Optimize Collaborative Dynamics Within the Global Film Production Network

Mehrdad Maghsoudi

*Department of Industrial and Information Management, Faculty of Management and Accounting, Shahid Beheshti University, Tehran, Iran*

Saeid Aliakbar

*Department of Industrial Engineering, Technical and Engineering Faculty of Damghan University, Semnan Province, Iran*

Sajjad HabibiPour

*Faculty of Mechanical Engineering, Khwaja Nasiruddin Tusi University of Technology, Iran*

## Abstract

Global film production is witnessing a surge in international cooperation and cross-border investment. However, the dynamics and structure underlying these collaborative ventures remain underexplored. This study employs social network analysis (SNA) to investigate joint investments within the cinema industry. A network of 150 countries is constructed based on 7800 links representing shared creative themes in film productions.

Centrality measures identify pivotal nodes like the United States, China, and England as influential countries with strong leadership potential to steer industry growth through collaboration. Community detection further uncovers distinct thematic clusters centered around common creative endeavors such as action films or social issues.

Each community reveals opportunities for targeted cooperation and investment. For instance, the "Global Thrill Seekers" community produces mainstream blockbusters, whereas the "Cultural-Social Cinema Group" tackles worldwide issues. Overall, this SNA enhances understanding of the global film network and highlights pathways to optimize data-driven decision-making regarding collaborative initiatives. The research underscores SNA's value in predicting outcomes and paving the way for strategic unity in the creative industries.

# 1. Introduction

The film industry has played a significant role in shaping and influencing various aspects of life and the current world (Vlassis, 2016). From its inception, the film has captivated audiences and provided a platform for storytelling, entertainment, and cultural expression(Vlassis, 2016; Vogel, 2020). Over the years, the film industry has evolved into a global phenomenon, attracting increased investments and fostering international cooperation(Dašić & Kostadinović, 2022).

In recent years, there has been a notable surge in investment within the film industry(Dastidar & Elliott, 2020). This trend can be attributed to several factors, including advancements in technology, growing demand for diverse content, and the increasing globalization of the film market(Betzler & Leuschen, 2021). As a result, stakeholders have recognized the immense potential and profitability of the film industry, leading to a rise in financial commitments from various entities(Behrens et al., 2021).

Furthermore, the film industry has witnessed a substantial increase in international cooperation and investments(Guo, 2023). Filmmakers, studios, and production companies are increasingly engaging in cross-border collaborations, joint ventures, and co-productions to leverage resources, expand market reach, and enhance creative synergies. This trend highlights the industry's recognition of the value in fostering global partnerships to create successful and impactful film experiences(Yayla et al., 2023).

However, despite the growing importance of collaborations and joint ventures in the film industry, there remains a need for comprehensive analysis, review, and forecasting in this domain. Quantitative and data-based analyses are crucial for understanding the dynamics, outcomes, and potential of these partnerships(Maghsoudi & Nezafati, 2023). Unfortunately, such analyses are often lacking, limiting the industry's ability to make informed decisions and optimize collaborative opportunities(Bulgurcu et al., 2018; Lee et al., 2020).

One powerful technique that can address this gap is social network analysis (SNA). Social network analysis is a methodological approach that examines the relationships and interactions between individuals, groups, organizations, or entities within a social system. By applying SNA, researchers can uncover hidden patterns, identify influential actors, analyze information flows, and predict outcomes within complex social, economic, and cultural networks(M. A. M. A. Kermani et al., 2022; Zohdi et al., 2022).

The aim of this paper is to utilize social network analysis in examining and evaluating collaboration opportunities and joint investments in the film industry. By employing SNA techniques, our objective is to elucidate the structure and dynamics of collaborative networks in the industry, identify key stakeholders and their roles, and provide insights for future decision-making..

The following sections of this article will include a literature review, where we explore existing research and theories related to collaborative ventures in the film industry. The methodology section will outline the approach and data sources used for our social network analysis. In the results section, we will present our findings and analysis based on the applied methodology. The discussion section will interpret the results, explore their implications, and discuss potential avenues for further research. Finally, the article will conclude with key insights, recommendations, and suggestions for policymakers, industry practitioners, and researchers interested in enhancing collaboration and predicting outcomes in the film industry.

2. **Literature review**
2.1. **Investments in the film industry**

Investments in the film industry play a vital role in the development and production of films, as well as the overall sustainability and growth of the industry(Johnsen, 2023; Messerlin & Parc, 2018). These investments can come from various sources, including studios, production companies, individual investors, and even crowdfunding platforms(Dastidar & Elliott, 2020). Understanding the different types of investments in the film industry and their potential returns is essential for both investors and industry professionals(Johnsen, 2023).

One common type of investment in the film industry is the financing of film production. Investors provide the necessary capital to fund various aspects of filmmaking, such as pre-production, production, post-production, marketing, and distribution. These investments can take the form of equity financing, where investors become shareholders in the film project, or debt financing, where investors provide loans to be repaid with interest(Johnsen, 2023).

Another type of investment in the film industry is through film funds or investment vehicles. Film funds pool together funds from multiple investors and allocate them to various film projects(Dastidar & Elliott, 2020). These funds are often managed by experienced professionals who assess the potential profitability and viability of film projects. Investors in film funds benefit from diversification, as their investments are spread across multiple projects, reducing the risk associated with investing in a single film(Johnsen, 2023).

Investments in the film industry can yield different types of returns. One primary return on investment is box office revenue(Dastidar & Elliott, 2020). When a film performs well at the box office, generating substantial ticket sales, investors can receive a share of the profits. However, box office success is not guaranteed, and films can also underperform, resulting in losses for investors(Johnsen, 2023).

In addition to box office revenue, films generate income from various revenue streams, including distribution deals, home video sales, streaming platforms, merchandise, and licensing. Investors may receive a portion of these revenues, depending on the terms of their investment agreements(Kübler et al., 2021). It is important to note that revenue-sharing models and profit participation can vary widely, depending on the specific agreements between investors and filmmakers(Johnsen, 2023).

Investments in the film industry also offer intangible returns, such as exposure and prestige. Successful films can elevate the profile and reputation of investors, opening doors to future investment opportunities and industry connections(Dastidar & Elliott, 2020). Furthermore, investments in the film industry provide individuals with the opportunity to support and contribute to the creation of meaningful and impactful storytelling(Johnsen, 2023).

However, it is crucial to acknowledge the inherent risks associated with film investments. The film industry is highly competitive, and not all projects achieve commercial success. Factors such as audience preferences, market conditions, critical reception, and marketing strategies can significantly impact the financial performance of a film. Therefore, investors must conduct thorough due diligence, assess the potential risks and rewards, and diversify their investment portfolios to mitigate risks(Johnsen, 2023).

## 2.2. Social Network Analysis (SNA)

Social Network Analysis (SNA) is a methodological approach that examines the relationships and interactions between individuals, groups, organizations, or entities within a social system. It provides a powerful framework for understanding the structure and dynamics of social networks and their influence on various phenomena. The origins of social network analysis can be traced back to the early 20th century, with notable contributions from sociologists, anthropologists, and mathematicians(HabibAgahi et al., 2022).

One of the pioneers of social network analysis was Jacob Moreno, who developed the sociogram in the 1930s. Moreno used this visual representation to depict social relationships and patterns among individuals within a group. This innovative approach laid the foundation for the formal study of social networks and their role in shaping human behavior and social dynamics(M. A. Kermani et al., 2022).

Social network analysis views events and phenomena through the lens of relationships and connections(Maghsoudi, Jalilvand Khosravi, et al., 2023). It recognizes that social structures and patterns of interaction can significantly impact the flow of information, influence, resources, and even the spread of behaviors or ideas(HabibAgahi et al., 2022). By analyzing the structure and properties of social networks, researchers can uncover hidden patterns, identify key actors or nodes, analyze information flows, and predict outcomes within complex social, economic, and cultural systems(Maghsoudi & Shumaly).

The main components of social network analysis are nodes, ties, and attributes. Nodes, also known as actors or vertices, represent individuals, groups, or organizations within the network. Ties, also referred to as edges or relationships, represent the connections or interactions between the nodes. These ties can be directed (one-way) or undirected (mutual)(Jalilvand Khosravi et al., 2022). Attributes refer to the characteristics or properties associated with nodes or ties, such as age, gender, profession, or the strength of a relationship(M. A. M. A. Kermani et al., 2022).

Social network analysis employs various indicators to analyze and measure the structure and properties of social networks. Three key indicators include degree, density, and centrality.

Degree is a basic measure of the number of connections or ties that a node has within a network. It provides an indication of an individual's popularity, influence, or connectedness within the network. The degree of a node can be calculated by counting the number of ties it has. In a directed network, the degree can be further classified as in-degree (number of incoming ties) and out-degree

(number of outgoing ties) to capture the flow of information or influence(Maghsoudi, Shokouhyar, Khanizadeh, et al., 2023).

Density is a measure of the interconnectedness or cohesion within a network. It quantifies the extent to which the nodes in a network are connected to one another. Density is calculated by dividing the actual number of ties present in the network by the total number of possible ties. Higher density values indicate a more interconnected network, while lower density values suggest a more fragmented or sparse network(Maghsoudi & Shumaly).

Centrality indicators measure the importance or prominence of nodes within a network. They identify nodes that occupy central positions and have a significant impact on information flow, influence, or communication. Some commonly used centrality indicators include degree centrality, betweenness centrality, and closeness centrality(M. A. Kermani et al., 2022; Maghsoudi, Shokouhyar, Ataei, et al., 2023).

- Degree centrality measures the number of ties that a node has relative to other nodes in the network. It provides an indication of an individual's popularity or influence within the network. Degree centrality can be calculated by dividing the number of ties connected to a node by the total number of nodes minus one.

$$C_d(v_i) = \sum_{i=1}^{n} a(v_i, v_j)$$

- Betweenness centrality captures the extent to which a node lies on the shortest paths between other nodes in the network. It measures the node's potential to control or facilitate the flow of information or resources between other nodes. Betweenness centrality can be calculated by determining the fraction of shortest paths in the network that pass through a particular node.

$$C_B(i) = \frac{1}{n^2} \sum_{\forall s,t \in V} \frac{n_{s,t}^i}{g_{s,t}}$$

- Closeness centrality reflects how quickly a node can access information or resources from other nodes in the network. It measures the average distance from a node to all other nodes in the network. Nodes with higher closeness centrality can disseminate information or exert influence more efficiently. Closeness centrality can be calculated by summing the shortest path distances from a node to all other nodes and dividing it by the total number of nodes minus one.

$$C_c(v) = \frac{n-1}{\sum_{i=1}^{n} d_{(v,i)}}$$

The concept of community in social network analysis refers to groups or subsets of nodes that are densely connected internally and sparsely connected externally. Communities represent clusters of nodes that exhibit stronger ties among themselves compared to nodes outside the community.

Detecting and understanding communities in a network can provide insights into the social structure, subgroups, and the flow of information or influence within the network(Maghsoudi, Shokouhyar, Khanizadeh, et al., 2023).

A community is a cluster of objects that exhibit closer relationships with each other than with other objects in the dataset. Within a community, individuals interact more frequently with fellow members than with those outside the group (Figure 1). The proximity of entities within a community can be evaluated by analyzing their similarity or distance from one another (Bedi & Sharma, 2016). In the context of a social network, a community can be likened to a cluster within the network (Said et al., 2018).

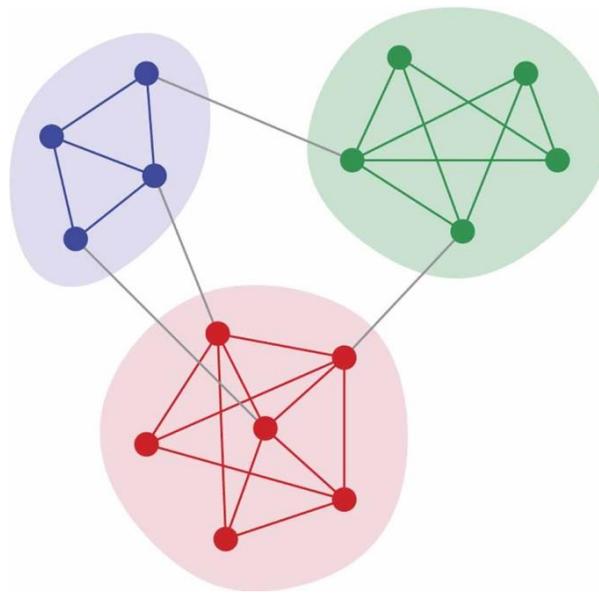

*Figure 1: Communities clustering (Ding et al., 2021)*

Community detection in network models involves creating community structures through clustering, as illustrated in Figure 2. According to the definition, a community is composed of nodes that have strong associations with one another, and these connections are stronger within the community than with nodes outside of it. Communities, representing groups of nodes with shared and similar properties, are valuable tools for network analysts seeking to comprehend interactions and cohesive sub-groups within the network (El-Moussaoui et al., 2019). Modularity, as defined by Newman, serves as a measure of social network clustering performance (Newman, 2006). Nodes that are interconnected exhibit a positive correlation with modularity.

In a weighted network with n nodes, the algorithm initiates by treating each node as an individual community. Thus, initially, there are multiple separate communities. The algorithm proceeds by evaluating each node (i) and identifying a neighboring community (j) that, when joined, maximizes the modularity index while removing node i from its current community. Node i is then added to community j if this change results in an increase in modularity; otherwise, node i remains in its original community. This process is iteratively repeated for all nodes until no further improvements can be made, leading to a locally optimal point in the first phase. This point represents a state where altering the communities of nodes does not yield any additional modularity gains.

In the second phase of the algorithm, the process continues by merging smaller communities that can be combined to form larger ones. These two phases persist until there are no more changes in the communities, and the modularity index reaches its maximum state.

As depicted in Figure 2, after completing the first phase and finding the local optimal point, the algorithm successfully identifies four distinct communities. In the second phase, it attempts to merge these four communities, eventually condensing them into two larger communities to achieve the highest possible modularity index, at which point the process comes to a halt.

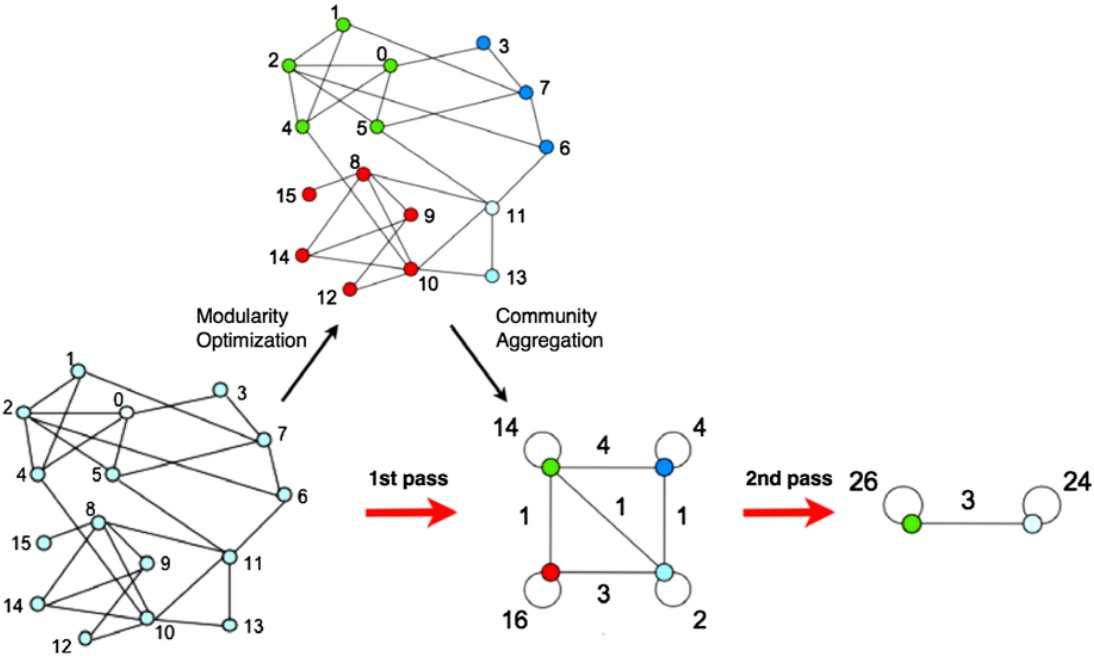

*Figure 2: Community detection based on increasing modularity(Maghsoudi, Jalilvand Khosravi, et al., 2023; Yap et al., 2019)*

## 2.6. Related Works

In the "Related Works" section of this manuscript, we explore a range of studies investigating different facets of the film industry. (Leem et al., 2023) present a 2023 study titled "Towards Data-Driven Decision-Making in the Korean Film Industry: An XAI Model for Box Office Analysis Using Dimension Reduction, Clustering, and Classification." This study centers on the growth of the Korean film market and the rising significance of explainable artificial intelligence (XAI) within the industry. To address the competitive nature of the market and the substantial expenses tied to film production, the authors propose the DRECE framework (Dimension REduction, Clustering, and classification for Explainable artificial intelligence). This framework employs dimensionality reduction techniques to transform complex data into a two-dimensional format, employs K-means clustering to group similar data points, and utilizes machine-learning models to classify movie clusters. Notably, the integration of XAI techniques ensures transparency in decision-making processes, offering insights that support film industry professionals in enhancing box office performance and optimizing profits. Adoption of the DRECE model holds the potential to provide fresh perspectives and insights, empowering decision-makers to strategically navigate the Korean film market for successful outcomes.

Another significant contribution comes from (Juhász et al., 2020), who delve into collaboration networks within the creative industries, with a specific focus on film production. This 2020 study examines the dynamics between creators occupying central and peripheral positions within these networks and how such roles impact creative success. While central creators benefit from prestige critical for creative achievement, they may lack exposure to diverse ideas originating from the periphery. The authors introduce the concept that central creators can enhance their prospects of creative success by acting as intermediaries, bridging the gap between the core and peripheral collaborators. This hypothesis is substantiated using a distinctive dataset encompassing Hungarian feature films from 1990 to 2009, dissecting the dynamic collaboration network among movie creators. The findings indicate that central creators who also serve as connectors between the core and periphery are notably more likely to achieve award-winning creative outcomes.

(Ding et al., 2022) contribute to the field through their study conducted in 2022, where they present the "WE model," a machine learning model tailored for predicting the derivatives market for films. The authors highlight that movie industry revenues stem not just from box office sales but also from merchandising, advertising, home entertainment, and other sources. Intriguingly, merchandising can even surpass box office sales in profitability. However, traditional market research techniques for forecasting merchandising markets across diverse film genres are labor-intensive and constrained. In response, the authors propose the WE model, which amalgamates three machine-learning algorithms to scrutinize crucial movie attributes and accurately forecast movie merchandising markets. By discerning the interplay between film attributes and merchandising markets, the WE model attains a 72.5% accuracy rate in predicting and assessing the derivatives market during testing. The study concludes that machine learning holds the potential to enable data-driven forecasting and management of movie merchandising markets, thereby demonstrating a successful application of machine learning techniques in predicting movie derivatives markets based on film attributes.

(Mateer, 2020) explores the landscape of academic-industry collaboration within the realm of commercial film and television production in the 2019 study titled "Academic-Industry Collaboration for Commercial Film and Television Production: an exploration of case studies." The study emphasizes the burgeoning collaborative models between academia and the film and television industry, aiming to enhance the practical relevance of academic programs while supporting the industry in identifying and nurturing new talent and optimizing production expenses. The paper evaluates two collaborative paradigms: the University as a "Production Partner" and the University as a "Service Provider." Drawing from diverse international collaborations, the study probes the structural aspects of these partnerships, the alignment of stakeholder needs, the benefits reaped by students and graduates, and the overall efficacy of such initiatives. A specific case study involving the University of York, UK, and Green Screen Productions Ltd. is also examined, underscoring the significance of alignment between engagement and institutional objectives for successful collaboration. The research posits that these collaborative models can be adapted to various global media contexts, contingent upon careful consideration of participants' needs and expectations.

(Ebbers & Wijnberg, 2012) contribute insights on the influence of reputation dimensions on distributor investments within the film industry. Their research discovers that a favorable reputation in a congruent dimension positively impacts investor behavior. Conversely, favorable reputations in incongruent dimensions weaken this positive effect, elucidating the phenomenon known as "reputational category spanning."

(Parc, 2018) investigates the repercussions of protectionist policies on the success of the Korean film industry. The study's findings indicate that protectionist measures such as import quotas and subsidies play a limited role in bolstering the industry's prosperity. Instead, market-friendly conditions and business activities emerge as pivotal factors in enhancing the industry's competitiveness.

(Powers, 2015) delves into the anticipated returns on tangible investments within the film industry. The study underscores the positive correlation between expected returns and the idiosyncratic dollar variance in a film's payoff. Additionally, the research explores the interrelationship between expected returns and dollar volatility, illustrating that heightened volatility corresponds to increased anticipated returns.

(Parc, 2021) turns the spotlight on the effects of business integration within the Korean film industry across different epochs. The study underscores the positive impact of a business-conducive environment during integration on the industry's competitive landscape. The study's findings hold significance for policymakers striving to shape effective cultural policies within the film industry.

(Zhang & Pelton, 2019) delve into the media representation of Wanda Group's investments in the U.S. entertainment sector amid the U.S.-China "trade war" during the Trump Administration. The research uncovers three core themes embedded in media coverage: business-related aspects, attitudes and actions of U.S. society, and China's soft power strategy. The study posits that the alignment of business endeavors with soft power strategy plays a pivotal role in shaping public sentiment and future relations between China and Hollywood.

(Kokas, 2020) sheds light on the investment dynamics within the Mainland Chinese film industry. The study illuminates the distinctive molecular structure characterizing national film investment, driven by the interplay between factors attracting commercial capital investment and those driving state centralization. This intricate investment pattern features diverse hubs interconnected by complementary bonds, shaped by regulatory frameworks, institutions, built environments, and access to capital.

Collectively, these works contribute significantly to the understanding of various aspects of the film industry, such as investment behavior, market conditions, and the influence of policy and reputation on its competitiveness. Researchers and policymakers can benefit from the insights provided by these studies to make informed decisions and foster growth in the film sector. Table 1 provides a summary of Related works.

*Table 1: summary of Related works*

| Year | Topic | Methodology | Key Results |
|---|---|---|---|
| 2023 | Data-Driven Decision-Making in the Korean Film Industry | XAI Model (DRECE Framework) with Dimension Reduction, Clustering, and Classification | DRECE Framework integrates XAI for transparent decision-making. Supports film professionals in enhancing box office performance. Empowers strategic navigation of the Korean film market. |
| 2020 | Collaboration Networks in Film Production | Network Analysis of Hungarian Feature Films (1990-2009) | Central creators acting as intermediaries enhance creative success. Bridging the gap between core and peripheral collaborators is beneficial. Central creators who connect core and periphery achieve award-winning outcomes. |
| 2022 | Predicting Derivatives Market for Films | "WE Model" with Machine Learning Algorithms | WE Model predicts movie merchandising markets accurately (72.5% accuracy). Machine learning helps forecast movie derivatives markets based on film attributes. Data-driven forecasting and management of merchandising markets are feasible. |
| 2020 | Academic-Industry Collaboration for Film and TV Production | Case Study Analysis of Academic-Industry Partnerships | Collaborative models enhance practical relevance of academic programs. Identify and nurture new talent in film and TV industry. Alignment between engagement and institutional objectives is crucial. |
| 2012 | Influence of Reputation Dimensions on Distributor Investments | Empirical Research on Reputation and Investment Behavior | Favorable reputation in congruent dimension positively impacts investor behavior. Favorable reputations in incongruent dimensions weaken this effect. Reputational category spanning" phenomenon explained. |
| 2018 | Protectionist Policies and Korean Film Industry | Analysis of Protectionist Measures and Industry Prosperity | Protectionist policies like import quotas and subsidies have limited impact. Market-friendly conditions and business activities are more influential |

| | | | Industry competitiveness is enhanced by business-friendly conditions. |
|---|---|---|---|
| 2015 | Returns on Tangible Investments in the Film Industry | Research on Expected Returns and Dollar Variance | Positive correlation between expected returns and film's payoff variance<br>Increased volatility corresponds to heightened anticipated returns<br>Expected returns linked to dollar volatility in film industry. |
| 2021 | Business Integration Effects in the Korean Film Industry | Study on Business Integration and Competitive Landscape | Positive impact of business-conducive environment during integration<br>Significant for shaping effective cultural policies in film industry<br>Business-friendly conditions enhance industry competitiveness. |
| 2019 | Media Representation of Wanda Group's Investments | Analysis of Media Coverage Amid U.S.-China Trade War | Three core themes in media coverage: business aspects, U.S. attitudes, China's soft power<br>Business alignment with soft power strategy shapes public sentiment<br>Media coverage influences China-Hollywood relations amid trade tensions. |
| 2020 | Investment Dynamics in Mainland Chinese Film Industry | Study on Investment Structure and Centralization | National film investment driven by factors attracting commercial capital<br>Interplay of factors driving state centralization shapes investment<br>Distinctive molecular structure of film investment driven by various factors. |

Our article builds on the existing body of research by analyzing and examining the cooperation and joint investments of different countries in the film industry through the technique of analyzing social networks and themes in films. This innovative approach differs from previous similar works as it focuses on the interconnectedness and collaborations between countries in the film sector, shedding light on how these relationships impact the industry's competitiveness and growth. By exploring social networks and themes, the article offers a unique perspective on the global dynamics of the film industry, providing valuable insights for both researchers and policymakers aiming to enhance international cooperation and investment in this sector.

## 3. Methodology

The steps of this research are based on Figure 3 and are as follows:

### 1. Data Collection:

The first step of our study involves collecting data related to movies, including their year of release, country of production, keywords, and other relevant information from the IMDB website. To efficiently gather this data, we employ the web crawling technique using Python programming language. The web crawler is programmed to extract movie data based on specified criteria such as the movie's production year and country of origin. The collected data will serve as the foundation for our subsequent analyses.

### 2. Matrix Creation:

Once the data collection process is completed, we proceed to create a communication matrix that represents the relationships between different countries in the film industry. This matrix is constructed using common keywords found in the films produced by various countries. The rows and columns of the matrix represent the names of the countries, while the matrix cells contain values indicating the number of shared keywords between the productions of the respective countries.

The fundamental principle underlying the matrix creation is that whenever two countries share a common keyword in their movie productions, we assign a value of one in the corresponding cell of the matrix. As a result, the more shared keywords two countries have, the higher their allocated number in the matrix will be. This communication matrix will provide valuable insights into the content affinity and potential cooperation between different countries in the film industry.

### 3. Network Formation:

Building upon the communication matrix from the previous step, we proceed to construct a communication network that visually represents the content affinity and cooperation potential between countries in the film industry. For the network visualization, we utilize the powerful Gephi software, widely recognized for its capacity in social network analysis. Gephi allows us to explore, analyze, and present the network in a visually appealing manner, making it an optimal choice for our research.

Using the communication matrix as input, Gephi generates a network graph, where countries are represented as nodes, and the connections (edges) between the nodes reflect the degree of shared keywords and content affinity between the respective countries. The network analysis will enable us to identify clusters or communities of countries that frequently collaborate or share similar themes in their film productions.

### 4. Network Analysis:

At this stage, we delve into analyzing the network at both micro and macro levels. At the micro level, we focus on the properties of individual nodes (countries) in the network, such as their degree centrality, betweenness centrality, and closeness centrality. These measures help us understand the importance of each country in terms of its connectivity, influence, and ability to act as a bridge between other countries in the network.

On the macro level, we examine global network properties, such as overall network density, Degree, and Edges. These metrics give us insights into the overall structure of the network and the presence of tightly knit communities or clusters within it.

As part of the network analysis, we employ community detection algorithms to identify cohesive groups of countries within the communication network. Community detection helps us identify clusters of countries that frequently collaborate or share thematic similarities in their film productions. Understanding these communities will shed light on the existing patterns of cooperation and joint investments among countries in the film industry.

**5. Policy Suggestions:**

Based on the findings from the previous steps, we draw meaningful conclusions about the cooperation and joint investments of different countries in the film industry. We present a set of policy suggestions and recommendations that can foster and strengthen international collaborations in filmmaking. These suggestions will be grounded in the empirical evidence obtained from our social network analysis and community detection, aiming to improve cross-border partnerships, boost the exchange of ideas, and encourage mutual investments to enhance the global film landscape.

By meticulously following these steps, our research provides a comprehensive and data-driven analysis of international cooperation in the film industry, utilizing techniques from social network analysis to unravel the intricate relationships and thematic patterns that shape the world of cinema.

*Figure 3: Research Methodology*

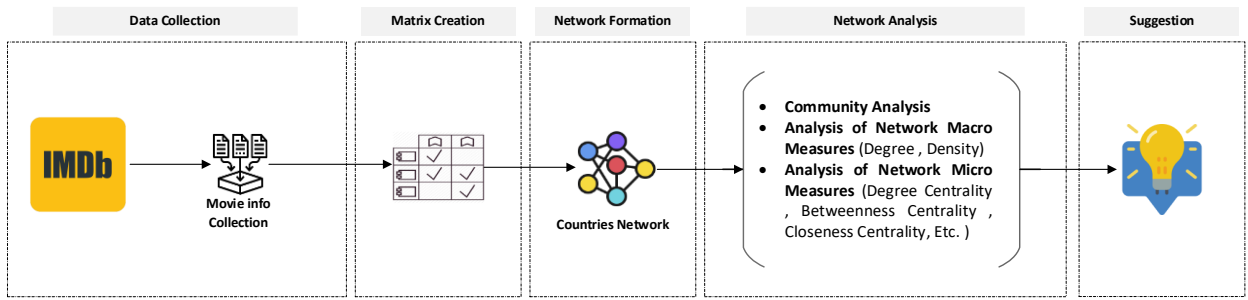

## 4. Results

### 4.1. Data Collection

In this study, movie data from the IMDB website was collected using Python web crawling libraries. Due to the extensive number of movies available on the platform, a filtering process was implemented to include only those movies with a score of more than 1000. As a result, a substantial dataset of 25,987 movies was successfully crawled. For a visual representation of the distribution of these crawled movies across different years, please refer to Figure 4.

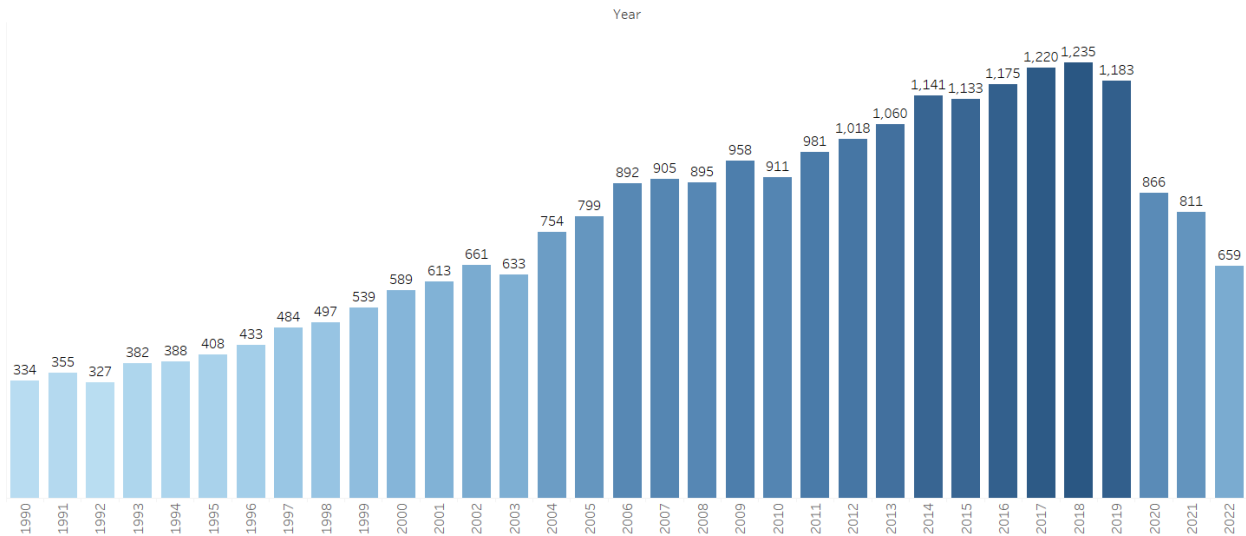

*Figure 4: Distribution of crawled movies by year of production*

During the keyword collection phase, a vast dataset comprising 3 million keywords was amassed. To enhance the precision and relevance of our analysis, a rigorous criterion was employed, focusing on keywords with a minimum of 3 user likes and fewer dislikes than likes. As a result of this stringent selection process, a total of 2.1 million unique keywords were retained for the subsequent stages of the study.

To further improve the relationship between keywords and facilitate a more comprehensive analysis, lemmatization was performed on this subset of unique keywords. Lemmatization is a text mining technique that involves transforming words to their base or root form, enabling us to group related words together and reduce the dimensionality of the dataset(Shumaly et al., 2021). This step resulted in a refined set of 38,000 keywords, which were thoroughly examined to explore meaningful relationships and patterns in the data.

### 4.2. Matrix Creation

In this phase of the research, we utilized the extracted keywords from movies to construct a matrix representing the relations between countries. The process involved cleaning and performing lemmatization on the extracted keywords. Subsequently, if a common keyword was found in two distinct movies, a connection (score) was established between the corresponding countries. The connection score between the countries was then incremented with each additional occurrence of this shared keyword in different movies.

For instance, if the keyword "violence" appeared in two films, one produced by the United States and the other by the United Kingdom, a connection was established between the United States and the United Kingdom, and a value of one was entered in the communication matrix. If this keyword recurred in 17 films from each country, the final number "17" was added to the matrix to indicate the strength of the connection between the two countries based on this keyword. This process was repeated for all relevant keywords, resulting in the creation of a comprehensive table similar to Figure 5, illustrating the interconnectedness between countries based on the presence of common keywords in their respective movies.

| Target | Arme.. | Aruba | Austr.. | Austria | Azerb.. | Baha.. | Bahra.. | Bangl.. | Barba.. |
|---|---|---|---|---|---|---|---|---|---|
| Cyprus | | | | | | | | | |
| Czech Re… | | | | 5 | | 1 | | | |
| Denmark | 3 | | | 7 | 1 | | | | |
| Dominica… | | | | | | | | | |
| Ecuador | | | | | | | | | |
| Egypt | | | | | 1 | | | | |
| Estonia | | | | | 1 | | | | |
| Finland | | | | 2 | 1 | | | | |
| France | 3 | 1 | 1 | 36 | 51 | | 1 | | 1 |
| Georgia | | | | | | | | | |
| Germany | ? | 1 | 2 | 36 | 90 | | 1 | | |
| Greece | | | | 2 | 2 | | | | |
| Guadelo.. | | | | | | | | | |
| Guatema.. | | | | | | | | | |
| Haiti | | | | | | | | | |
| Hong Ko.. | | | 1 | 4 | | | | | |
| Hungary | | | | 3 | 10 | | | | |
| Iceland | | | | 1 | | | | | |
| India | ? | | | 12 | 3 | | | | 7 |
| Indonesia | | | | 2 | | | | | |
| Iran | | | | 2 | 2 | | | | |
| Iraq | | | | | 1 | | | | |

*Figure 5: An example of the created matrix*

## 4.3. Network Formation

In the previous phase, a matrix was generated to establish and visualize the communication network among countries, with a specific focus on shared concepts in film productions. The outcomes of this phase are portrayed in Figure 6. The network incorporates 150 countries, interconnected through a total of 7,800 links between country nodes, reflecting an average degree of 52. This emphasizes a notable degree of collaboration and convergence in thematic elements across cinematic works. The network demonstrates consistency and unity, as indicated by a density of 0.35, thus highlighting a substantial level of connectivity within the global film industry. Furthermore, the presence of a single connected component within the network, encompassing all 150 countries, signifies a cohesive and integrated network structure.

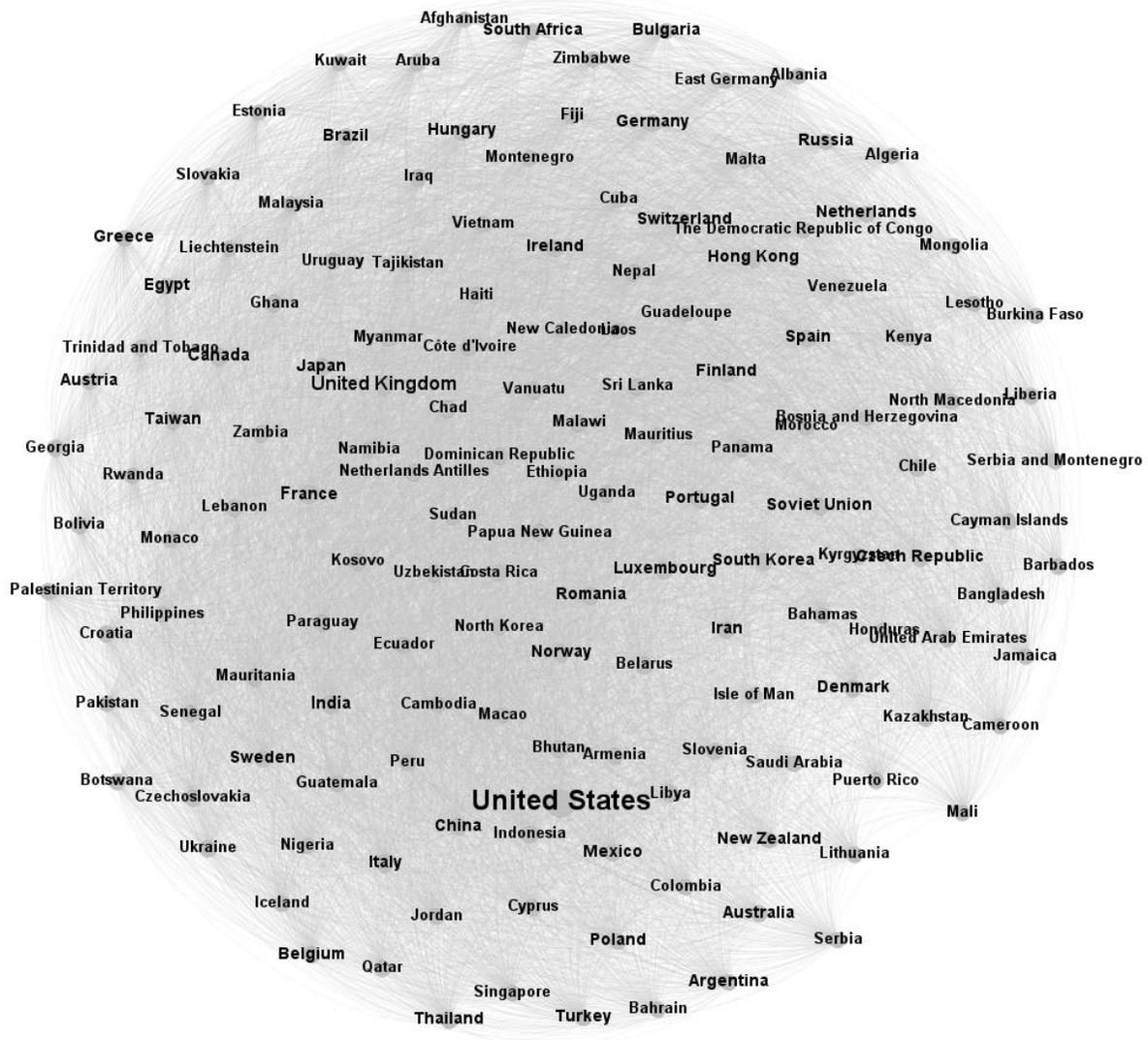

*Figure 6: Conceptual network of countries*

## 4.4. Network Analysis:

## 4.4.2. Network Micro Measures:

Applying centrality measures provides a powerful technique for micro-level examination of networks. These metrics identify the most significant and influential components in a network based on different criteria. As shown in Table 2, the top 20 countries ranked by degree, betweenness, closeness, and eigenvalue centrality reveal the nations that hold critical positions and substantial influence within the network. Such countries serve pivotal roles and strongly shape the dynamics of the overall network. Focusing further analysis on these high centrality countries will offer crucial insights into the underlying mechanisms and relationships governing the network. The centrality metrics spotlight the key nodes that deserve greater attention when seeking to grasp the nuances of the broader collaborative ecosystem.

*Table 2: top 20 in centrality Measures*

| Degree centrality | Betweenness centrality | Closeness centrality | Eigen centrality |
|---|---|---|---|
| United States | United States | United States | United States |
| United Kingdom | China | China | China |
| France | United Kingdom | United Kingdom | United Kingdom |
| Germany | Japan | Japan | Japan |
| Japan | France | France | France |
| China | India | India | India |
| Canada | Germany | Germany | Germany |
| India | South Korea | South Korea | South Korea |
| South Korea | Canada | Canada | Canada |
| Italy | Australia | Australia | Australia |
| Spain | Italy | Italy | Italy |
| Australia | Spain | Spain | Spain |
| Brazil | Mexico | Mexico | Mexico |
| Mexico | Hong Kong | Hong Kong | Hong Kong |
| Russia | Belgium | Belgium | Belgium |
| Sweden | Netherlands | Netherlands | Netherlands |
| Poland | Poland | Norway | Poland |
| Belgium | Hungary | Denmark | Sweden |
| Denmark | Denmark | Ireland | Brazil |
| Netherlands | Ireland | Hungary | Russia |

analyzing the centrality indices in this network significant insights into the cooperation and investment capacities of different nations.

### Degree centrality

Degree centrality measures the number of connections a node (country) has in the network. In this case, it represents the number of common themes shared between countries' films. The United States, with the highest degree centrality, implies it has collaborated or shared common film themes with a significant number of countries. This could be due to its large film industry and influence on global cinema trends.

### Betweenness centrality

Betweenness centrality identifies countries that act as bridges or intermediaries between other countries in the network. The United States, once again, holds the highest betweenness centrality, suggesting it plays a crucial role in connecting other countries due to its extensive collaborations. China and the United Kingdom also have high betweenness centrality, indicating their influence in facilitating connections.

### Closeness centrality

Closeness centrality measures how quickly a country can reach other countries in the network. The United States, with the highest closeness centrality, indicates that it has relatively short paths to reach other countries through shared film themes. China and the United Kingdom also have high closeness centrality, reflecting their efficient access to other countries.

**Eigenvector centrality**

Eigenvector centrality considers not only a country's connections but also the quality of its connections (i.e., how connected its connections are). The United States has the highest eigenvector centrality, indicating that it is connected to other influential countries. China, the United Kingdom, and Japan also score high in eigenvector centrality, suggesting they have connections to countries that themselves have many connections.

The countries with the highest centrality indices, including the United States, China, the United Kingdom, France, and Japan, showcase their prominence and influence in the global film production network. These countries are not only prolific film producers but also actively collaborate and share common themes with a diverse range of other countries.

The United States consistently ranks at the top in all centrality measures, reinforcing its status as a global film production hub. It has numerous connections, acts as a bridge between countries, is efficiently reachable, and has quality connections.

China's high centrality indices indicate its increasing importance in the global film industry. Its influence in bridging gaps between other countries is noteworthy. The United Kingdom's centrality emphasizes its historical ties to global cinema and its continuing role in connecting different film markets. France and Japan's strong centrality scores underline their cultural significance and their ability to engage with a broad spectrum of nations.

These high centrality index countries not only contribute significantly to the film production landscape but also serve as key players in fostering international collaboration and investment in the industry. Their centrality underscores their potential to shape trends, share expertise, and drive innovation in global cinema.

### 4.4.3 Community Detection and Analysis:

Community detection is a powerful analytical tool that allows researchers to identify coherent groups in a network based on the connections between its nodes. By applying community detection algorithms to the network, we can discover distinct clusters of countries with similar creative endeavors and thematic approaches in their filmmaking. By identifying these clusters, we can gain a deeper understanding of the connectivity and potential for collaboration in the global film production industry. Figure 7 shows a representation of the result of discovering associations in the network.

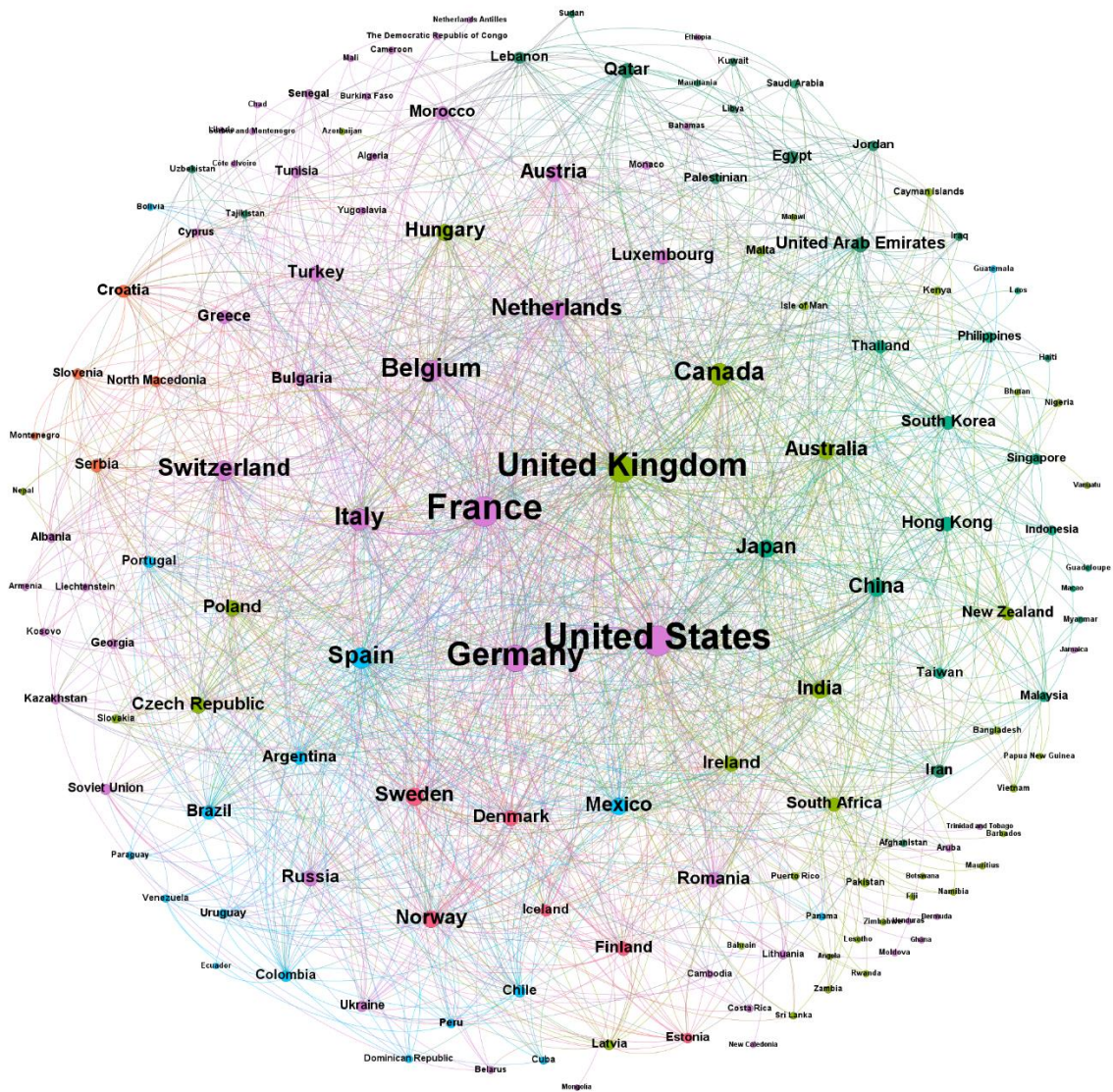

*Figure 7: Communication network of countries after community detection*

Table 3 shows the details of the countries present in each forum and the most frequent concepts in the products of that forum.

*Table 3: The main countries present in each community along with the most frequent keywords*

| color | Nodes | edges | density | The most important countries | Keywords |
|---|---|---|---|---|---|
| 🟩 | 39 | 107 | 0.14 | England, Canada, India, Australia, Hungary, Poland, Ireland, Czech Republic, South Africa, New Zealand | Confrontation, Gunfight, Near Death Experience, Apology, Surrealism, Interrogation, Anger, Forest |
| 🟪 | 53 | 280 | 0.20 | United States, France, Germany, Belgium, Italy, | Politics, police, anger, street market, Gun fight, Cemetery, |

| | | | | Switzerland, Netherlands, Austria, Russia, Turkey | Night club, bound and gagged, lawyer, revolver |
|---|---|---|---|---|---|
| | 16 | 42 | 0.35 | Qatar, United Arab Emirates, Iran, Egypt, Lebanon, Palestine, Jordan, Saudi Arabia, Iraq, Kuwait | prayer, religion, ambush, tent, survival, poverty, desert, terrorism, corruption, politics, jail, tragic event |
| | 5 | 9 | 0.9 | Croatia, Serbia, North Macedonia, Slovenia | Prisoner, comedy, cheating wife, priest, unfaithfulness |
| | 17 | 54 | 0.39 | Spain, Mexico, Brazil, Argentina, Chile, Portugal, Colombia, Uruguay, Peru | Cemetery, Near Death Experience, Apology, Ambush, Pries, Urination, Thief, Surrealism, Hallucination |
| | 6 | 13 | 0.86 | Sweden, Norway, Denmark, Finland, Iceland | Underwater Scene, Surrealism, Jungle, rural setting, extramarital affair, guilt, apology, forest |
| | 15 | 47 | 0.44 | Japan, China, Hong Kong, South Korea, Thailand, Taiwan, Philippines, Singapore, Malaysia | urban environment, ambush, warrior, bow and arrow, surrealism, massacre, gun fight, near death experience |

Based on the countries and keywords in each community, we can name and analyze each community as follows:

**Community 0: Global Thrill Seekers**

The prevalence of words like "confrontation," "gunfight," "near death experience," and "anger" suggest these film industries frequently produce action and dramatic films that involve high stakes, danger, and emotional intensity. This may reflect cultural interests in these countries for exciting, gripping stories. However, the inclusion of more cerebral words like "surrealism" indicates these industries also produce thought-provoking, avant-garde films that break conventions.

The appearance of "apology" and "interrogation" implies many films involve moral dilemmas, conflicts, and the complexity of human nature. There is an interest in exploring redemption, justice, truth, and reconciliation. Meanwhile, words like "forest" denote an appreciation for nature and the environment in cinematic storytelling.

Taken together, these keywords paint a portrait of diverse, developed film industries producing well-rounded content - from mainstream blockbusters to art house films. The range of stories suggests strong creative talent able to make films with both commercial success and critical acclaim.

For investors, this highlights opportunities to back films with potential international breakout appeal if they achieve the right balance of mass entertainment and artistic merit. These film industries possess the infrastructure and skill to make globally resonant movies, although adequate funding and marketing remains imperative. Overall, there is significant upside in these film markets if investors can identify the standout content and provide the necessary production and promotion support.

**Community 1: Socio-Political Cinema**

The concepts suggest films from these countries often deal with crime, violence, and darker themes. Politics and anger indicate social commentary and tensions. The settings like street markets, nightclubs, and cemeteries paint an urban landscape.

This reflects the diversity of stories told through film in these countries - from hard-hitting dramas and thrillers to nuanced character studies. The prominence of guns and revolvers shows action is a popular genre. Bound and gagged implies these films do not shy away from grittier subject matter.

For investors, this highlights the range of films with potential crossover appeal. There is clearly an audience for well-executed genres like thriller, drama, and action. The films also act as a lens into the cultures and social issues of each country. This insight helps identify themes and talents that can translate globally.

Additionally, the concepts suggest established filmmaking infrastructures in these countries. The ability to tackle complex narratives indicates expertise across production roles like directing, writing, cinematography. This talent can be leveraged for international co-productions.

In summary, these concepts reveal seasoned film industries adept at tackling challenging stories. For investors, it signals opportunities in supporting films and filmmakers that reflect the diversity of human experiences. This allows connection with worldwide audiences looking for authentic perspectives.

**Community 2: Faith and Unity**

The frequent appearance of words like prayer, religion, desert, and tent point to the cultural and geographical context that shapes many Middle Eastern films. Religion is deeply ingrained in society in most of these countries, and the desert landscape is a dominant visual motif. This reflects the spiritual values and harsh environmental realities that locals navigate daily.

However, words like terrorism, corruption, politics, jail, poverty, and tragic event also emerge often. This suggests that Middle Eastern cinema frequently depicts sociopolitical issues and harsh realities faced by citizens. It grapples with topics like government oppression, violence, and economic struggle. This reveals some mainstream perceptions of the region as turbulent.

Ambush also comes up, which may refer to scenes of military conflict. However, this could provide a complex look into the nuances of war from different perspectives, rather than simplistic portrayals of violence.

Overall, these keywords indicate that Middle Eastern cinema offers unique windows into the cultural fabric, visual landscapes, and real-life issues of the region. For investors, this presents an opportunity to fund projects that authentically capture unheard narratives. It allows global audiences to gain deeper understanding of the identities and stories of people from these countries through artful cinematic storytelling. The rich setting and drama inherent in these topics gives strong creative material.

By supporting Middle Eastern filmmakers in telling their own stories with nuance, investors can gain both financial returns and the chance to increase cultural awareness. The region's complex realities offer compelling content that distills both the breathtaking beauty and humanity of daily life in these countries. There is universal resonance in such storytelling.

**Community 3: Balkan Connections**

The prevalence of words like "prisoner", "cheating wife", and "unfaithfulness" suggests these film industries often explore darker themes related to crime, punishment, and moral transgressions. This indicates a cultural interest in examining the darker sides of human nature and relationships through film. The frequent appearance of "priest" also suggests an engagement with themes of morality, sin, and redemption.

However, the high frequency of "comedy" indicates these film industries also have a strong tradition of lighthearted, humorous films that aim to entertain and make audiences laugh. The combination of these concepts points to filmmakers willing to explore the full range of human experiences and emotions.

For investors, this analysis highlights both an audience appetite for serious dramas about social issues, as well as comedies that provide escapist entertainment. The diversity of genres and tones in these regional film industries provides opportunities to finance films tailored to local tastes that can also break out to wider international audiences. Overall, the concepts indicate strong local traditions of storytelling that connect with audiences, which are attractive features for potential investors.

The analysis shows these regional film industries have carved out artistic identities that sets them apart in the broader landscape of European and world cinema. By financing films that authentically capture the cultural sensibilities of the Balkan region, investors can tap into established local creative talent and interest while co-producing films with potential for global reach.

**Community 4: Latin Elegy**

The prevalence of words like "cemetery", "near death experience", "ambush" and "hallucination" indicates that films from these countries often explore dark, surreal and macabre themes. This reflects a strong tradition of magical realism and avant-garde storytelling that distinguishes Latin American cinema in particular. Directors like Alejandro Jodorowsky, Guillermo del Toro and Pedro Almodóvar have brought these unconventional narratives to international audiences.

Themes of death and violence further speak to the social realities experienced in many of these countries. Crime, political instability and inequality have been ongoing struggles, leading to reflective art forms. Yet there is also an apology, alluding to themes of redemption, forgiveness and hope.

The mention of thieves and urination introduces an earthy, gritty realism to the films' styles. They don't shy away from the rawness and vulgarity of everyday life. This makes them highly relatable to local audiences, though potentially controversial to more socially conservative nations.

For investors, this analysis highlights the creative boldness and insightfulness that defines much of Spanish and Latin American cinema. There is strong demand for new perspectives in global entertainment, making these film industries viable options. Their productions require modest budgets compared to Hollywood, allowing investors' resources to go further. But their exotic locations, talented crews and connection with local communities offer valuable comparative advantages. They are close to the cultural pulse of an influential region. With the right financing and distribution partnerships, their films have significant potential to cross over to worldwide audiences.

In summary, these conceptual keywords paint Spanish and Latin American films as imaginative, socially-conscious and intimately local. For investors, they offer the right mix of creativity and realism to potentially achieve both critical and commercial success in the global marketplace. Their uniqueness commands attention amidst a crowded entertainment industry.

**Community 5: Nordic Odyssey**

The prevalence of surrealism suggests Nordic films often explore the blurring of fantasy and reality. Directors like Ingmar Bergman used dreamlike sequences and existential symbolism to create psychological tapestries. This penchant for the abstract provides opportunities for avant-garde cinema that challenges audiences. Investors may want to back auteurs with singular visions.

Underwater scenes and rural settings indicate a deep connection to the natural environment. Nordic storytelling often uses the mysteries of nature to explore the human condition. The loneliness of a forest or the vastness of the sea can mirror inner emotional landscapes. Financiers could target films that use the tangible wilderness as a gateway to intangible feelings.

Guilt and apologies demonstrate a focus on morality and the conscience. Nordic dramas tend to wrestle with the gray areas of ethics and responsibility. The prevalence of extramarital affairs in particular reveals an interest in the tension between passion and social contracts. Investors may find promise in complex character studies that foreground moral dilemmas.

While Hollywood blockbusters emphasize spectacle, Nordic films seem to prize nuance, ambiguity, and atmosphere. Supporting auteur-driven projects with unconventional narratives could help distinguish Nordic cinema in global markets. Patient financing that allows for bold artistic risk-taking is key. Overall, these conceptual patterns suggest strong potential to establish a distinctive regional voice.

**Community 6: Asian Enigma**

The prevalence of words like "urban environment," "ambush," "warrior," "bow and arrow," "surrealism," "massacre," "gun fight," and "near death experience" in films from these countries highlights a few key themes:

Action and Violence - There is clearly a strong emphasis on action, combat, and depictions of violence in many popular films from these regions. The references to warriors, ambushes, massacres, and gun fights point to thrilling action sequences and intense fight scenes being major draws for audiences.

Historical Settings - The mentions of warriors, bows and arrows, and surrealism suggest many films are set in historical or period settings. There is an appreciation for classics tales and historical legends retold on film with modern effects and stylization.

Dark/Gritty Tone - The violence, action, and near-death experiences evoke a gritty, somber, or dark tone in many of these films. They point to films that do not shy away from visceral violence, high stakes, and mortal peril for the characters.

For investors and producers looking at these film industries, it highlights some key considerations:

Action-centric films with highly choreographed fight/battle scenes tend to draw big crowds and buzz in these markets. Investing in these types of films can pay off financially.

Classic folklore, legends, and historical periods are rich sources of inspiration that connect with audiences who value their cultural heritage.

While Hollywood goes for more family-friendly superhero films, there is a contrasting appetite in these regions for edgier, grittier, more violent action pics.

Big spectacle films built around combat, feats of human ability, and impressive CG visuals cater towards the visual engagement and excitement that draws audiences.

Known stories and characters adapted into live-action films appeal to built-in fanbases, as long as the adaptation feels authentic and respectful.

Focusing on these action, violence, and culturally-rooted themes can be a key to success for investors seeking to capitalize on the lucrative Asia Pacific film market. The analysis shows a divergence from Hollywood towards grittier, more visceral and culturally inspired films.

# 5. Conclusion

This study conducted an analysis of the communication network among 150 countries based on common themes found in their film productions. The findings of this research revealed a highly interconnected global network with substantial cooperation and convergence in cinematic concepts.

The network analysis identified countries with the highest influence, as demonstrated by their high centrality scores. The United States, China, and England emerged as key players, reflecting their prominence in driving and pioneering innovative trends in the industry. Their centrality underscores their leadership in fostering international collaboration and investment.

The identification of communities unveiled distinct clusters of countries united by common creative approaches. For instance, the "Global Thrill Seekers" community produces action films, while the "Socio-Political Cinema Society" delves into social issues. Each community offers opportunities for targeted collaboration and investment based on shared interests.

Broadly, this research highlights contributions and growth potential in the global film production landscape. As national cinemas increasingly converge through collaborative investments, joint productions, and shared creative themes, new prospects for strategic unity and investment initiatives emerge. Regionally appropriate budgets, collaborative production agreements, film festivals, and network platforms can all contribute to enhancing these connections.

Utilizing network analysis and community identification techniques, stakeholders can make informed decisions regarding focal points for collaboration and investment to maximize impact. Just as the network evolves over time, further analysis can identify growing capacities and emerging creative communities.

**Policy Recommendations:**

- Drawing insights from social network analysis and the article's findings, here are some policy recommendations to enhance collaborative efforts and investments in the global film industry:
- Establish international joint production funds to financially support film projects that align with creative or thematic interests shared between countries. Such funds can facilitate the creation of high-quality, globally appealing productions through shared resources. Targeted budgets could be allocated for identified communities like "Global Thrill Seekers" or "Faith and Unity."
- Initiate cultural exchange programs to promote the exchange of talents, ideas, and best practices among countries through workshops, educational programs, residencies, etc. This approach can enrich skills and perspectives.
- Launch region-specific film funds dedicated to projects supporting cultural heritage and narratives in specific regions like the Middle East, Northern Europe, the Balkans, Latin America, etc. This approach encourages local collaboration.
- Organize international film festivals with a focus on outstanding productions from identified communities. These events provide platforms for showcasing films, networking, and discussions.

- Develop joint marketing and distribution strategies to promote films from a community globally. This enhances accessibility and influence through coordinated efforts.
- Standardize policies for collaborative production, incentive schemes, and treaty frameworks to enable seamless cross-border cooperation between countries.
- Establish joint training programs between national film institutions to develop specialized skills necessary for international productions.
- Create databases and platforms for sharing information about best practices in collaborative production, success stories, upcoming projects, etc., to enhance accessibility.
- Generate incentives for private investment funds specifically geared towards financing international joint productions, especially from key identified countries. This incentivizes new capital resources.

In conclusion, this study's comprehensive analysis sheds light on the potential of global film production collaborations and growth. As the boundaries between national cinemas blur, fostering strategic alliances and investment becomes paramount. The proposed policy recommendations aim to guide stakeholders in harnessing the power of collaborative creativity to shape the future of the global film industry.